\documentstyle[12pt,citation]{article}

\font\tenrm=cmr10
\font\tenit=cmti10
\font\elevenbf=cmbx10 scaled\magstep 1
\font\elevenrm=cmr10 scaled\magstep 1
\font\elevenit=cmti10 scaled\magstep 1

\font\ninerm=cmr9

\def\agt{\buildrel {\mbox{$>$}} \over {\raisebox{-0.8ex}{\hspace{-0.05in}
$\sim$}}}
\def\alt{\buildrel {\mbox{$<$}} \over {\raisebox{-0.8ex}{\hspace{-0.05in}
$\sim$}}}
\renewenvironment{thebibliography}[1]
 { \elevenrm
   \begin{list}{\arabic{enumi}.}
    {\usecounter{enumi} \setlength{\parsep}{0pt}
     \setlength{\itemsep}{3pt} \settowidth{\labelwidth}{#1.}
     \sloppy
    }}{\end{list}}

\begin{document}
\hfill\vbox{\hbox{\bf NUHEP-TH-93-15}\hbox{Jun 1993}}\par
\begin{center}{{\bf Standard Model Higgs Boson at Linear Photon Colliders
\footnote{\ninerm Talk given at the ``Workshop on Physics and Experiments with
Linear $e^+e^-$ Collider", Waikoloa, Kona, Hawaii}
\\}
\vglue 5pt
{\rm Kingman Cheung\\}
\baselineskip=13pt
{\tenit Department of Physics, Northwestern University, \\}
\baselineskip=12pt
{\tenit Evantson, Illinois  60208, U.S.A.\\}
\vglue 0.3cm
{\tenrm ABSTRACT}}
\end{center}
\vglue 0.3cm
{\rightskip=3pc
 \leftskip=3pc
 \tenrm\baselineskip=12pt
 \noindent
We will summarize some aspects of the standard model Higgs boson at high
energy
$\gamma\gamma$ colliders, where the photon beams are realized by
the laser backscattering method, including  the direct Higgs boson
production via $\gamma\gamma\rightarrow H$, and the measurement of the Yukawa
coupling via the channel $\gamma\gamma\rightarrow t\bar tH$.
%
\vglue 0.6cm}
\baselineskip=14pt

For the past few years considerable amount of efforts have
 been exploring on the
physical possibilities at high energy $\gamma\gamma$ and $e\gamma$ collisions
\cite{bord}, with the photon beams converted from the electron or positron
beams in linear $e^+e^-$ colliders by laser backscattering method
\cite{teln}.  One of
these possibilities is the study of the standard model (SM) Higgs boson.
Despite  the difficulties in the measurement of the photon-Higgs coupling in
the Higgs boson decay, the direct Higgs production via
$\gamma\gamma\rightarrow H$ provides a feasible measurement of the coupling.
We will show the feasibility of using the decay channel of
$\gamma\gamma\rightarrow H\rightarrow
ZZ\rightarrow q\bar q \ell\bar \ell$ in searching for a heavy Higgs boson with
$m_H
\agt 200$~GeV at a 0.5~TeV $e^+e^-$ collider.  This decay mode should be clean
except it suffers from the continuum productions of $ZZ$ via box diagrams,
$q\bar q Z$, $\ell\bar \ell Z$, and $t\bar t \rightarrow b\bar b \ell\bar
\ell\nu\bar \nu$.

Another interesting process, the associated  production of
$\gamma\gamma\rightarrow t\bar tH$,
provides a good opportunity to study the Yukawa-top coupling
for an intermediate mass Higgs (IMH) boson at  1--2 TeV $e^+e^-$ colliders.
  In this
channel we will confine to the IMH boson because
the event rate is too low for heavy Higgs boson.  We will also compare this
process with $e^+e^-\rightarrow t\bar tH$ \cite{djo}.

\vglue 0.6cm
{\elevenbf \noindent 1. Laser Backscattering}
\vglue 0.4cm

Hard photon beams at $e^+e^-$ machines can be produced by
directing a low energy (a few $eV$) laser beam at a very small angle
$\alpha_0$, almost head to head, to the incident electron beam.
Through Compton scattering, there are abundant, hard back-scattered
photons in the same direction as the incident electron.
The energy spectrum of the unpolarized
back-scattered photon is given by \cite{teln}
\begin{equation}
\label{lum}
F_{\gamma /e}(x) = \frac{1}{D(\xi)} \left[ 1-x +\frac{1}{1-x}
-\frac{4x}{\xi(1-x)} + \frac{4x^2}{\xi^2 (1-x)^2} \right] \,,
\end{equation}
where
\begin{equation}
\label{D_xi}
D(\xi) = (1-\frac{4}{\xi} -\frac{8}{\xi^2}) \ln(1+\xi) + \frac{1}{2} +
\frac{8}{\xi} - \frac{1}{2(1+\xi)^2}\,,
\end{equation}
$\xi=4E_0\omega_0/m_e^2$, $\omega_0$ is the energy of the incident laser
photon, $x=\omega/E_0$ is the fraction of the incident electron's
energy carried by the back-scattered photon, and the maximum value $x_{\rm
max}$ is given by  $x_{\rm max}= \xi/(1+\xi)$.
\vglue 0.6cm
{\elevenbf \noindent 2. $\gamma\gamma\rightarrow H\rightarrow ZZ$
\cite{bowser}}
\vglue 0.4cm

One of the Feynman diagrams is  depicted in Fig.~\ref{feyn1}.
All massive charged particles contribute to the loop,
in particular, this process could be used to probe the Higgs sector with  a
heavy top quark \cite{boos}, or to detect the presence of new ultra-heavy
fermions \cite{chano} or non-SM $W'$ charged gauge bosons. In this report,
we assume only SM particles in the triangular loop; in particular, we
have used a top-quark mass of 150~GeV.

\begin{figure}[t]
\vspace{2in}
\caption{\label{feyn1}
\tenrm \baselineskip=12pt
 Feynman diagram for $\gamma\gamma\rightarrow
H\rightarrow ZZ$.}
\end{figure}

\baselineskip=14pt
The hard scattering cross section for $\gamma\gamma \to H\to Z Z$, for
monochromatic photons with $m(\gamma\gamma)=\sqrt{\hat s}$, is given by:
\begin{equation}
\label{sigmahat}
\hat\sigma(\hat s) =
 {{8\pi \Gamma(H^*\to\gamma\gamma)\Gamma(H^*\to Z Z)}
        \over
  {(\hat s - m_H^2)^2 + \Gamma_H^2 m_H^2}}\label{hard}\,,
\end{equation}
where the partial widths in the numerator are those of a virtual Higgs of
mass $m_{H^*}=\sqrt{\hat s}$. The on-resonance cross section $\sigma_0$ is
simply given by $\sigma_0 = \hat\sigma(m_H^2)$. We have calculated the
total cross section $\sigma(\gamma\gamma\to H\to ZZ)$ by folding in
Eq.~(\ref{sigmahat}) with the photon luminosity function of
Eq.~(\ref{lum}).  These cross sections $\hat \sigma$ and $\sigma$ are shown in
Fig.~\ref{cross}.

For the range of $m_H$ considered here ($m_H \alt
400$~GeV), the true width $\Gamma_H$ is quite small. Then the partial widths
can be approximated by their on-shell values, obtaining:
\begin{equation}
\label{sigma}
\sigma(s) = \sigma_0 \,\,\, \left(\frac{\pi m_H \Gamma_H}{s}\right)
         \int^{x_{\rm max}}_{m_H^2/(x_{\rm max}s)}
         F_{\gamma/e}(x_1) F_{\gamma/e}(\tau/x_1) \frac{dx_1}{x_1}\,,
\end{equation}
where $\tau=m_H^2/s$ and $\sqrt{s}$ is the center-of-mass energy of the
parent $e^+e^-$ collider. From Eq.~(\ref{sigma})  we anticipate that the
total cross section should roughly decrease with increasing collider
energy $\sqrt{s}$.

\begin{figure}[t]
\vspace{3in}
\caption{\label{cross}
\tenrm \baselineskip=12pt
(a) The resonance cross section $\hat \sigma$, (b) the actual cross
section $\sigma$ folded with the photon luminosity function, for the process
$\gamma\gamma\rightarrow H\rightarrow ZZ$.}
\end{figure}

\baselineskip=14pt
The non-monochromatic nature of the photon beams
drastically reduces the Higgs production cross section from its on-resonance
value. The ``gold-plated'' detection mode $ZZ \to
\ell^+\ell^-\ell^+\ell^-$  is thus marginal
because of the small branching fractions.
On the other hand, $ZZ\rightarrow q\bar q q'\bar q',\, q\bar q \nu\bar \nu,\,
\ell\bar \ell \nu\bar \nu$ are  all
obscured by the huge continuum background from $\gamma\gamma\to WW$.  We
finally choose the mixed hadronic and leptonic decay mode $ZZ\rightarrow q\bar
q \ell\bar \ell$.  To eliminate the $WW$ background we require all the quarks
and leptons to be observable  in the final state, with
\begin{equation}
\label{coscut}
 z = \max\left\{|\cos\theta_i|\right\} < 0.95\,,
\qquad i=q,\bar q, \ell^+ \ell^-
\end{equation}
Although this decay mode is rather clean, it does suffer from
$q\bar qZ,\, \ell\bar \ell Z, t\bar t \rightarrow b\bar b
\ell\bar \ell\nu\bar \nu$ induced backgrounds.
The other background $\gamma\gamma\rightarrow ZZ$
via box diagrams \cite{jikia}, which we did not include here,
turns out not negligible, but it
will not affect our conclusion here for $m_H$ upto 300 GeV.
To estimate these backgrounds we impose a constraint of the $Z$ mass on the
$q\bar q$ and $\ell\bar \ell$ pairs
\begin{equation}
\label{mass}
 |m(q\bar q\,{\rm or}\,\ell^+\ell^-) - m_Z| < 8\;{\rm GeV}\,,
\end{equation}
and the $t\bar t$ background turns out negligible after this cut.
The other backgrounds from $q\bar qZ$ and $\ell\bar\ell Z$ tend to be
 very forward/backward, and the cut in
Eq.~(\ref{coscut})
can effectively reduce these backgrounds, as shown in Fig.~\ref{direct}a.
We show our results on the $m(ZZ)$ spectrum after the acceptance cuts
of Eqs.~(\ref{coscut}) and (\ref{mass}) in Fig.~\ref{direct}b.

{}From Fig.~\ref{direct}b we can see that the Higgs-boson peaks are rather
distinct for $m_H=200-300$~GeV, but less sharp for $m_H\agt300$~GeV. To
estimate the feasibility of discovery we define the signal $S$ and background
$B$ as  the integrated cross section in the interval
\begin{equation}
\label{mzzcut}
m_H\pm \Gamma\,,\qquad  {\rm where}\; \Gamma={\rm max}(\Gamma_{\rm
resolution},\, \Gamma_H)\,,
\end{equation}
in the $m(ZZ)$ spectrum,
and we take a rather conservative $\Gamma_{\rm resolution}=10$~GeV.  We
present the signal $S$, background $B$, and the significance of the signal
$S/\sqrt{B}$ in Table~\ref{table1} assuming an integrated luminosity of
20~fb$^{-1}$.

\begin{figure}[t]
\vspace{3in}
\caption{\label{direct}
\tenrm \baselineskip=12pt
(a) The integrated cross sections as a function of $z_{\rm cut}$ for
the signal and the sum of the backgrounds (dashed),
(b) the differential cross section
$d\sigma/dm(ZZ)$ for the signal of various Higgs masses  and the sum of
the backgrounds after the acceptance cuts.}
\end{figure}

\begin{table}[b]
\caption{\label{table1}
Table showing the cross section for the signal $S$,
background $B$, and the significance of the signal for $m_H$=200--400 GeV.}
\centering
 \baselineskip=14pt
\begin{tabular}{|c|ccc|}
\hline
$m_H$   &  $S$(fb)  &  $B$(fb)    &  $S/\sqrt{B}$ (20 fb$^{-1}$)  \\
\hline
200     &  1.6  & 0.34    &  12   \\
250     &  1.4  & 0.25    & 12  \\
300     &  0.92 & 0.14    & 11 \\
350     & 0.38  & 0.13    &  4.7 \\
400     & 0.099 & 0.075   &  1.6 \\
\hline
\end{tabular}
\end{table}

In conclusion of this part, the discovery of the heavy Higgs boson
for $m_H=200-300$ GeV should be viable at the 0.5 TeV $e^+e^-$ collider
operating in the $\gamma\gamma$ mode.   Increase in center-of-mass energy will
not improve the feasibility as indicated by Eq.~(\ref{sigma}).

\vglue 0.6cm
{\elevenbf\noindent 3. $\gamma\gamma\rightarrow t\bar tH$ \cite{ttH}}
\vglue 0.4cm

At tree level the  Higgs-boson  couples to a fermion of
mass $m_f$ with the strength $g_{ffH} \sim  g m_f/2m_W$.
The coupling  $g_{ffH}$ can be directly probed in the decay of the
Higgs boson into a pair of fermions if kinematically allowed.
To measure $g_{ttH}$ directly by this method however we need $m_H > 2 m_t$.
Therefore,  this method cannot be used to probe the $g_{ttH}$ coupling
directly for the intermediate mass Higgs boson ($m_W<m_H < 2m_Z$).
It can be probed indirectly  in the decay of $H\rightarrow
\gamma\gamma$ or $gg$, or in the
fusion of $\gamma\gamma\rightarrow H$ or
$gg\rightarrow H$
through an internal  top-quark loop; but it is likely to be
affected by the presence of other heavy particles beyond the SM.
An alternative  direct probe is to use the associated production of a
Higgs boson  with a $t\bar t$ pair at the $e^+e^-$ colliders \cite{djo}.
In principle, the same coupling can also be probed  in the production process
 $e^+e^- \rightarrow t\bar tZ$ \cite{hagi}, but the contribution  from
the Higgs-exchange diagram  is very small
for the Higgs mass below the $t\bar t$ threshold.
Therefore, in the case of IMH $t\bar tZ$ production cannot
probe the Higgs-top coupling, but becomes a potential
background to $t\bar tH$ production, especially if $m_H$ is close to
$m_Z$.

\begin{figure}[t]
\vspace{3in}
\caption{\label{tthfig}
\tenrm \baselineskip=12pt
Total cross sections versus center-of-mass energies of the parent
$e^+e^-$ collider, for $m_H$=90 GeV and $m_t$=150 GeV.
The subprocesses $\gamma\gamma \rightarrow t\bar tH$ (solid), $t\bar tZ$
(dashed); $e^+e^- \rightarrow t\bar tH$ (dotted), $t\bar tZ$
(dash-dotted) are shown.}
\end{figure}

\baselineskip=14pt

The process  $\gamma\gamma \rightarrow t \bar t H$
offers another possible direct test of the $g_{ttH}$ coupling in addition
to $t\bar tH$ production in $e^+e^-$ collisions.
The signature, due to the dominant
decays of    $H \rightarrow b\bar b$ and $t \rightarrow bW$, is
\begin{equation}
\label{bbbbWW}
\gamma \gamma \rightarrow t\bar t H\rightarrow b\bar b b\bar b WW\, .
\end{equation}
Almost all the backgrounds can be removed by using the constraints
due to the  $W$, $t$ and $H$ masses.
Even so, $t\bar tZ$ production is a potential
background, especially if the Higgs mass is close to the $Z$ mass.
In the following we will show that
the process $\gamma\gamma\rightarrow t\bar tH$, in the light of the
$t\bar tZ$ background, is better than the process $e^+e^-\rightarrow t\bar tH$
in probing the Yukawa-top coupling.

In Fig.~\ref{tthfig} we show  the total cross sections versus the
center-of-mass energies $\sqrt{s}$ of the parent $e^+e^-$ collider for
$m_H=90$~GeV and $m_t$=150~GeV.
For $e^+e^-$ collisions  both
$t\bar tH$ and $t\bar tZ$ productions reach a maximum between
$\sqrt{s}=500$~GeV to 750 GeV, and then fall gradually as $\sqrt{s}$
increases further, which is due to
the $s$-channel  $\gamma$ or $Z$ propagator.
Roughly, $t\bar tZ$ production in $e^+e^-$ collisions is about a
factor of 2 to 5 larger than production of $t\bar tH$.  Consequently, if
$m_H$ is close to $m_Z$ the $t\bar tH$ signal could be difficult to identify
due to the potential $t\bar tZ$ background.

On the other hand, the cross sections for
$\gamma\gamma \rightarrow t\bar tH$ and $t\bar tZ$ start off very
small at $\sqrt{s}=500$~GeV because only a very narrow range of $x$ is
available at this energy.
But both increase gradually as $\sqrt{s}$ increases, because a growing
range of $x$ is available and there is no propagator contributing
a factor $1/s$ as in the corresponding case of $e^+e^-$ collisions.
For $m_t$=150 GeV  $t\bar tH$
production is larger than $t\bar tZ$ production for the whole range of
energies shown in Fig.~\ref{tthfig}.
This is an important advantage of $\gamma\gamma$
collisions over $e^+e^-$ collisions for probing the $g_{ttH}$
Yukawa coupling.

For $\sqrt{s}$ from 0.5 TeV to about 1~TeV  the $e^+e^-\rightarrow t\bar tH$
cross section (1.5--3~fb) are larger than the
$\gamma\gamma\rightarrow t\bar tH$ cross sections. However, for this
range of $\sqrt{s}$, the potential background from $t\bar tZ$ production
(3--6 fb) is also larger in $e^+e^-$ collisions.
For $\sqrt{s}\agt 1$~TeV $\gamma\gamma$ collisions provide a better
approach than $e^+e^-$ collisions since the cross section ($\sim
1.5-2$~fb) is larger and there is less potential background from
$t\bar tZ$ production ($\sim 0.5-2.5$~fb).

\vglue 0.6cm
{\elevenbf \noindent 4. Conclusions}
\vglue 0.4cm

High energy $\gamma\gamma$ colliders provide good opportunities to search for
the SM Higgs boson upto about 300 GeV at a 0.5 TeV $e^+e^-$ machine, and
$\gamma\gamma\rightarrow t\bar tH$
turns out to be a better channel to measure the Yukawa
coupling than the process $e^+e^-\rightarrow t\bar tH$ for $\sqrt{s_{ee}}\agt
1$ TeV.

\vglue 0.6cm
{\elevenbf \noindent Acknowledgements:}
Most of the work presented here was in collaboration with David Bowser-Chao.
Special thanks to the organizers of the workshop for financial support,
and to S.~Kanda for his hospitality during the workshop.
This work was supported in part by the DOE grant DE-FG02-91-ER40684.

\vglue 0.6cm
{\elevenbf \noindent Reference}
\vglue 0.4cm

\end{document}